\documentclass[fleqn,usenatbib,useAMS]{mnras}

\usepackage[T1]{fontenc}

\DeclareRobustCommand{\VAN}[3]{#2}
\let\VANthebibliography\thebibliography
\def\thebibliography{\DeclareRobustCommand{\VAN}[3]{##3}\VANthebibliography}

\usepackage{graphicx}	
\usepackage{amsmath}	
\usepackage{amssymb}	
\usepackage[utf8]{inputenc}
\usepackage{xcolor}
\usepackage{bm}

\usepackage{newtxtext,newtxmath}

\allowdisplaybreaks   

\numberwithin{equation}{section}

\newcommand{\FT}[1]{\mathcal{F} \{ #1\}} 
\def\bi#1{\hbox{\boldmath{$#1$}}}

\title[Self-Calibrating the Look-Elsewhere Effect]{
Self-Calibrating the Look-Elsewhere Effect: \\
Fast Evaluation of the Statistical Significance Using Peak Heights
}

\author[A.~E.~Bayer, U.~Seljak and J.~Robnik]{
Adrian E.~Bayer,$^{1,2}$\thanks{E-mail: abayer@berkeley.edu}
Uro\v{s} Seljak$^{1,2,3}$\thanks{E-mail: useljak@berkeley.edu}
and
Jakob Robnik$^{4}$\thanks{E-mail: jrobnik@student.ethz.ch} 
\\
$^{1}$Berkeley Center for Cosmological Physics, University of California, Berkeley, CA 94720, USA\\
$^{2}$Department of Physics, University of California, Berkeley, CA 94720, USA\\
$^{3}$Physics Division, Lawrence Berkeley National Laboratory, Berkeley, CA 94720, USA\\
$^{4}$Department of Physics, ETH-H\"onggerberg, 8093 Z\"urich, Switzerland\\
}

\date{Accepted XXX. Received YYY; in original form ZZZ}

\pubyear{2021}

\begin{document}
\label{firstpage}
\pagerange{\pageref{firstpage}--\pageref{lastpage}}
\maketitle

\begin{abstract}
In experiments where one searches a large parameter space for an anomaly, one often finds many spurious noise-induced peaks in the likelihood. This is known as the look-elsewhere effect, and must be corrected for when performing statistical analysis.  
This paper introduces a method to calibrate the false alarm probability (FAP), or $p$-value, for a given dataset by considering the heights of the highest peaks in the likelihood.
In the simplest form of self-calibration, the look-elsewhere-corrected $\chi^2$ of a physical peak is approximated by the $\chi^2$ of the peak minus the $\chi^2$ of the highest noise-induced peak. Generalizing this concept to consider lower peaks provides a fast method to quantify the statistical significance with improved accuracy.
In contrast to alternative methods, this approach has negligible computational cost as peaks in the likelihood are a byproduct of every peak-search analysis.
We apply to examples from astronomy, including planet detection, periodograms, and cosmology.
\end{abstract}



\section{Introduction}

When searching a large parameter space for a signal, with limited a priori knowledge of the signal's location, the likelihood distribution is often multimodal with the vast majority of peaks corresponding to spurious noise-induced events.
This is known as the look-elsewhere effect, or problem of multiple comparisons, and must be accounted for when performing a hypothesis test to avoid reporting a false detection \citep{Miller_1981, mult_hyp_test}. 

This effect is particularly prevalent in astronomy, with numerous examples including 
searching 
for gravitational waves and exoplanets.
In the gravity wave example, one searches for a signal of many different possible known shapes and unknown time, which can lead to a large look-elsewhere effect, modifying the $p$-value by many orders of magnitude (see e.g.~\cite{cannon2015likelihoodratio, Abbott_2016, Messick_2017}).
A similarly large effect occurs when searching for exoplanetary transits in
stellar photometry 
experiments such as Kepler \citep{kepler}, where
the period, phase, and other properties of the transit are unknown \citep{Baluev_2008,Baluev_2013,Baluev_2015,Delisle_2020}).
Additionally, the look-elsewhere effect can occur in wavelet analysis: this has many astronomical applications, one of which is detecting asteroid families in the main belt \citep[see e.g.][and references therein]{Baluev_2018,Baluev_2018b,Baluev_2020,Baluev_2020b}.

The look-elsewhere effect is also prominent in searches for new particles, where the mass of the particle is unknown: this issue gained much attention when the LHC detected the Higgs boson \citep{Aad:2012tfa, Chatrchyan:2012xdj}.
Just like collider searches, astroparticle experiments also suffer from the look-elsewhere effect, with examples including:
constraining the dark matter self-annihilation cross-section via
gamma ray emission from galaxy clusters \citep{Anderson_2016},
searching for
WIMPs via charged cosmic rays \citep{Reinert_2018}, searching for non-baryonic dark matter via X-ray emission from the Milky Way \citep{10.1093/pasj/psv081}, explaining the source of high energy astrophysical neutrinos \citep{Aartsen_2014, Emig_2015}, and in the spectral analysis of solar neutrinos \citep{Ranucci_2007}.
The look-elsewhere effect additionally appears in inflationary cosmology when searching for anomalies
in the primordial power spectrum \citep{Fergusson:2014hya,Fergusson:2014tza,Hunt_2015}, and when 
detecting planar structures in the satellite systems of galaxies 
\citep{10.1093/mnras/stv1557}.

The look-elsewhere effect is also relevant in numerous areas outside of physics. In biology, modern DNA sampling techniques can be used to perform genetic association to find links between genotypes and phenotypes \citep{Genes1, Genes2}. When large DNA sequences are used there is a high probability of obtaining spurious signals and thus a large look-elsewhere effect. Another medical example is the process of testing the effectiveness of drugs in clinical trials \citep{drugs}. Furthermore, it is important to consider when attempting to find hidden prophecies in ancient religious texts \citep{BibleCode2_mckay1999}.
%
%
The look-elsewhere effect is ubiquitous in physics and beyond and there is thus much motivation for a fast method to account for it.

When performing a hypothesis test,
frequentists typically consider the $p$-value, whereas Bayesians consider the Bayes factor. The $p$-value is often referred to as the false positive rate (FPR) or false alarm probability (FAP), as it quantifies how often a given test statistic is expected to take on a particular, or more extreme, value under the assumptions of the null hypothesis. Hence, the smaller the $p$-value, the less likely the null hypothesis and the larger the statistical significance of the alternative
hypothesis. 
Typically one considers the $p$-value of the likelihood ratio between the alternative and the 
null hypotheses, in which case the 
look-elsewhere effect causes an increase in the $p$-value at a fixed value of the likelihood: discretely speaking, if one performs $N$ trials, the probability of a spurious 
event increases by a factor of $N$. Conversely, the likelihood required to achieve a given $p$-value is increased by the look-elsewhere effect, meaning one needs to find peaks with a larger likelihood to achieve a given statistical significance. 

A brute force method to account for the change in the $p$-value is to perform simulations of the null hypothesis and determine the $p$-value numerically. This, however, is extremely computationally expensive: for example, to determine the 5-sigma level, which corresponds to a $p$-value of $\sim 10^{-7}$, one would need to perform $\sim 10^7$ simulations.
To achieve a balance between efficiency and accuracy, scientists have often applied the theory of \cite{Davies, Davies2} to find an upper bound for the $p$-value. \cite{Baluev_2008, Baluev_2013, Baluev_2015} applied this to the Lomb-Scargle periodogram \citep{Lomb_1976,Scargle_1982}, and various generalizations, by performing intricate analytical calculations. However, these calculations depend on the type of periodogram considered and are thus not applicable to more general situations. For example, in the task considered by \cite{Baluev_2021} it was found that analytic approximations are either inaccurate or slow to compute.  Furthermore, \cite{Gross} applied the theory of \cite{Davies, Davies2} in the context of particle searches by using the expected number of upcrossings to approximate the asymptotic $p$-value via the Taylor approximation. This method still requires multiple simulations, although fewer than the brute force approach.

An alternative 
approach to account for the look-elsewhere effect, which requires neither simulations nor model-specific calculation, 
has been recently developed by \cite{LEE1}.
Drawing a connection with Bayesian methodology, it associates the look-elsewhere effect \textit{trials factor} with the 
prior-to-posterior volume ratio.
This method has been shown to be effective for a variety of models when evaluating the Bayes factor using the Laplace approximation and the posterior volume using the Hessian matrix.
This makes the computation very fast, however, when considering models of increased complexity where the Laplace
approximation is not valid, 
one may have to employ Monte Carlo (MC) methods (e.g.~\cite{Chen_2000}) to accurately evaluate the Bayes factor, which would increase the computational time.

In this paper we work towards a general approach to account for the look-elsewhere effect, which requires neither simulation, nor model-specific calculations, nor explicit evaluation of the Bayes factor.
To achieve this, we consider the distribution of peak heights in the multimodal likelihood computed from the data. 
We show that, given this information alone, one can estimate the $p$-value directly from the likelihood of the data in examples of varying complexity. 
Since the peak heights are a byproduct of a
peak search, this information is readily available and provides a very fast way to estimate the $p$-value. 
Because this approach accounts for the look-elsewhere effect by using information from the data alone, we name the method \textit{self-calibration}.

The paper is organized as follows. Section \ref{sec:theory} reviews the 
look-elsewhere effect
and defines quantities such as the trials factor. Section \ref{sec:selfcalib} then motivates the method to 
self-calibrate the $p$-value and trials factor. This is then illustrated in Section \ref{sec:results} for a variety of examples related to planet detection (spots in exoplanetary transit light curves, the LS periodogram, and a Kepler exoplanet search) and an example from cosmology (searching for oscillatory features in the primordial power spectrum).
Finally, conclusions are given in Section \ref{sec:conclusions}.

\section{Background}
\label{sec:theory}

This section briefly summarizes the look-elsewhere effect (see \cite{LEE1} for a more thorough discussion).
We consider a model with $M$ parameters, $\bi{z}$, such that $z_1$ is the amplitude of a signal, and $\bi{z}_{>1}$ describes the properties of the signal that one is scanning over, or fitting for. 
For example, searching for the signal of a planet transit with amplitude $z_1$, orbital period $z_2$, and phase $z_3$.

From a frequentist perspective, 
one is interested in comparing the data $\bi{x}$ against the null hypothesis $H_0$ of there being no signal ($z_1=0$). Writing the likelihood as $p(\bi{x}|\bi{z})$, a common test statistic to consider is related to the likelihood ratio,
\begin{equation}
    q_L(\bi{z}) \equiv 2 \ln \frac{p(\bi{x}|\bi{z})}{p(\bi{x}|\bi{z}_0)},
    \label{qL}
\end{equation}    
where $\bi{z}_0$ represents the values of the parameters under $H_0$. 
To assess the significance one typically 
considers the maximum value of $q_L$, denoted $\hat{q}_L$, but below 
we will generalize this concept.
For a Gaussian likelihood, $q_L$ is equal to the difference in $\chi^2$ between the null and signal hypotheses, thus we will often simply refer to this as the chi-squared. In such a case, and in the absence of the look-elsewhere effect, $\sqrt{\hat{q}_L}$ gives the number-of-sigma significance. This can then be related to the $p$-value, often referred to as the false positive rate (FPR) or false alarm probability (FAP), depending on the problem in question: for example in the case of a chi-squared random variable with $s$ degrees of freedom, the $p$-value is given by the complementary cumulative distribution function of a chi-squared random variable with $s$ degrees of freedom.
The $p$-value in the absence of the look-elsewhere effected is referred to as the \textit{local} $p$-value.

In the presence of the look-elsewhere effect the $p$-value must be corrected, sometimes by many orders of magnitude. This is often done by introducing the trials factor, $N$, such that
the look-elsewhere-corrected $p$-value
is parameterized as
\begin{equation}
    P(\hat{Q}_L > \hat{q}_L) = N P_{\rm local}(\hat{Q}_L > \hat{q}_L),
    \label{pql_global}
\end{equation}
where $P_{\rm local}$ is the local $p$-value, and $\hat{Q}_L$ is the random variable associated with $\hat{q}_L$. The look-elsewhere-corrected $p$-value is often referred to as the \textit{global} $p$-value as it considers the probability that the global maximum of the likelihood occurs above a specific threshold. 
The trials factor thus quantifies the extent of the look-elsewhere effect, with $N=1$ corresponding to no look-elsewhere effect, and progressively larger values corresponding to a more severe look-elsewhere effect. Computing the trials factor is the main challenge in accounting for the look-elsewhere effect, and is typically done either by (i) performing model-dependent analytical calculations \citep{Baluev_2008,Baluev_2013,Baluev_2015}, (ii) running many numerical simulations \citep{Gross},  or (iii) by evaluating the Bayes factor \citep{LEE1}. In the next section we present self-calibration as a fast method to estimate $N$ and the $p$-value directly from the data, without any expensive computation.

\section{The Self-calibration Method}
\label{sec:selfcalib}

In this section we introduce our proposal for self-calibrating the $p$-value using the heights of peaks in the likelihood. We present a Bayesian derivation in Appendix \ref{app:selfcalib}, but give a more concise frequentist motivation in this section.

In the presence of a large look-elsewhere effect ($N \gg 1$), the parameter space will contain many peaks. Various works have derived the number of upcrossings (or the number of local maxima in the case of multidimensional fields) that breach a particular threshold, $\tau$, of the chi-squared \citep[see e.g.][]{Rice_1945, Adler_1981, Davies, Davies2, Azais_Delmas_2002}.
One can write the expected number of upcrossings as
\begin{equation}
    \langle n_{\rm up}(\tau) \rangle = C \tau^\alpha e^{-\tau/2},
    \label{eq:n_up}
\end{equation}
where $\alpha$ and $C$ are model dependent quantities, and we have ignored boundary effects which are negligible for large trials factor $N$. Note that the trials factor is absorbed into $C$. 

By evaluating the expected number of upcrossings from Eq.~\ref{eq:n_up} at the maximum chi-squared value, denoted $\hat{q}_L$, one obtains an asymptotic (large $\hat{q}_L$) approximation of the $p$-value as 
\begin{equation}
    P(\hat{Q}_L \geq \hat{q}_L) \simeq \langle n_{\rm up}(\hat{q}_L) \rangle = C \hat{q}_L^\alpha e^{-\hat{q}_L/2}.
    \label{eq:p_up}
\end{equation}
In the case of chi-squared random variable with $s$ degrees of freedom $\alpha=(s-1)/2$ \citep{Davies2}. Many problems of physical interest obey this form, for example a periodogram with 1 harmonic signal corresponds to $s=2$ \citep{Baluev_2008}, or $s=2h$ in the case of $h$ harmonic signals \citep{Baluev_2009}, while particle physicists hunting for a mass resonance typically consider $s=1$ \citep{Cowan_2011}. We will assume this form of $\alpha$ in the remainder of the text.\footnote{One can consider non-linear models for which $\tau^\alpha$ is replaced by a model-dependent polynomial function of $\tau$ \citep[see e.g.][]{Baluev_2013}, however we will demonstrate that this is often unnecessary to a good approximation in §\ref{sec:kepler}.}

On the other hand, the coefficient $C$ can be complicated to compute as it is sensitive to the look-elsewhere effect and has a high degree of model dependence. Efforts have been made to compute it for many different scenarios \citep{Baluev_2008, Baluev_2009, Baluev_2013, Baluev_2015}, however, this requires a specific calculation for each example considered, and for many problems can become analytically intractable \citep{Baluev_2021}. 

To avoid having to compute $C$, we eliminate it by combining Eqs.~\ref{eq:n_up} and \ref{eq:p_up}. This gives
\begin{align}
    P(\hat{Q}_L>\hat{q}_L) 
    &\simeq \langle n_{\rm up}(\tau) \rangle \left( \frac{\hat{q}_L}{\tau} \right)^{(s-1)/2} e^{-(\hat{q}_L - \tau)/2} \\
    &= e^{ -\frac{1}{2} \left[\hat{q}_L-\tau-2 \ln \langle n_{\rm up}(\tau) \rangle - (s-1) \ln \frac{\hat{q}_L}{\tau} \right] }.
    \label{eq:gross}
\end{align}
In essence, Eq.~\ref{eq:gross} is calibrating the expected number of upcrossings at the maximum value of the chi-squared, $\hat{q}_L$, by using the number of upcrossings at a lower value from the chi-squared, $\tau$.

This relation in Eq.~\ref{eq:gross} was also found by \cite{Gross} using a slightly different argument. In order to apply their method, \cite{Gross} suggests to evaluate $\langle n_{\rm up}(\tau) \rangle$ by performing thousands of simulations and computing the numerical average. While this requires fewer simulations than evaluating the $p$-value directly it can still be computationally expensive \citep{Algeri_2016}.

In self-calibration we bypass such simulations, and quickly obtain the $p$-value directly from a single dataset. We do this by firstly
noting that the number of peaks is a good approximation to the number of upcrossings in the asymptotic $\hat{q}_L$ limit. Using peaks instead of upcrossings is beneficial because peaks are a byproduct of any peak-search analysis, making them readily available.  Secondly, we use the fact that, by definition, there are $n$ peaks with $q_L$ larger than or equal to the $n^{\rm th}$ highest peak, $q_L^{(n)}$. We thus take $\tau \approx q_L^{(n)}$ and $\langle n_{\rm up} (\tau) \rangle \approx n$.
Substituting this into Eq.~\ref{eq:gross} enables application to a single dataset, negating the need for simulations.
Finally, we apply the \v{S}id\'{a}k correction \citep{sidak} to Eq.~\ref{eq:gross} to improve the non-asymptotic behaviour. The \v{S}id\'{a}k correction replaces the asymptotic $p$-value such that $P \rightarrow 1 - e^{-P}$, to provide better agreement in the non-asymptotic (large $P$) regime \cite[see e.g.][]{LEE1}.

In this paper we will consider models with 1 amplitude parameter and $s$ other parameters, i.e. $s$ degrees of freedom, thus the total number of parameters is given by $M=s+1$.
Using bars to denote self-calibrated values, the  self-calibrated $p$-value is thus given by 
\begin{equation}
\bar{P}(\hat{Q}_L>\hat{q}_L) \equiv 1-\exp \left(-e^{ -\frac{1}{2} \left[\hat{q}_L-\tau_n-2 \ln n - (M-2) \ln \frac{\hat{q}_L}{\tau_n} \right] } \right),
 \label{sidak_selfcalib}
\end{equation}
where $\tau_n$ is known as the \textit{threshold}. We provide discussion on choices of $\tau_n$ in Appendix \ref{app:selfcalib}, but $\tau_n \approx q_L^{(n)}$ is a suitable approximation which we will often employ. Hence, Eq.~\ref{sidak_selfcalib} relates the $p$-value to an exponential function of the difference in peak height between the highest peak and the $n^{\rm th}$ highest peak (via the $\hat{q}_L-\tau_n\approx\hat{q}_L-q_L^{(n)}$ term), and some logarithmic correction terms that depend on the choice of $n$ and the dimensionality of the model $M$. We will discuss the choice of $n$ in Section \ref{sec:results}.

The form of the self-calibrated calibrated $p$-value can be more succinctly written as
\begin{equation}
\bar{P}(\hat{Q}_S>\hat{q}_S) \equiv 1-\exp \left(-e^{ -\bar{\hat{q}}_S/2} \right),
 \label{sidak_selfcalib_qS}
\end{equation}
where
\begin{equation}
    \bar{\hat{q}}_S \equiv \hat{q}_L-\tau_n-2 \ln n - (M-2) \ln \frac{\hat{q}_L}{\tau_n}.
    \label{qap}
\end{equation}
It is beneficial to work with the $q_S$ test statistic instead of $q_L$ as the look-elsewhere dependence is absorbed into $q_S$, making the $p$-value relation in Eq.~\ref{sidak_selfcalib_qS} independent of the look-elsewhere effect \citep{LEE1}.
In the case of a one-tailed test (i.e.~positive amplitude), it follows from \cite{LEE1} that the self-calibrated statistical significance, or number of sigma, $\bar{S}$, is given by
\begin{align}
    \bar{S} &= \sqrt{\bar{\hat{q}}_S - \ln 2 \pi \bar{\hat{q}}_S},
    \label{S_selfcalib}
\end{align}
for large $\bar{\hat{q}}_S$. Note for sufficiently large $\bar{\hat{q}}_S$, $\bar{S} \approx \sqrt{\bar{\hat{q}}_S}$.
Moreover, the self-calibrated trials factor, $\bar{N}$, is given by
\begin{align}
  2 \ln \bar{N}
   = \tau_n + 2 \ln n + \ln 2 \pi \hat{q}_L + (M-2) \ln \frac{\hat{q}_L}{\tau_n}.
  \label{tlnNsc}
\end{align}

The above discussion concerns the distribution of peak heights for pure noise. 
In practice, a dataset might contain one, or more, physical peaks; applying Eq.~\ref{sidak_selfcalib} to such a dataset would give the significance of a signal under the null hypothesis, i.e.~assuming all peaks are noise. The existence of physical peaks will
cause an overestimation of $\tau_n$ and in turn an overestimation of the $p$-value, or an underestimation of the significance. This could introduce false negatives, but not false positives, making this a conservative approach.
However, this overestimation of the $p$-value will be small for $n$ larger than a few due to the slowly varying logarithmic term $\ln n$. 
In cases where there are multiple peaks introduced by physical sources, one can reduce this effect by iteratively removing non-maximal physical signals and appropriately relabeling the peaks in terms of $n$.
More generally, Eq.~\ref{qap} shows that to self-calibrate $\hat{q}_S$ for $M=2$ one must correct $\hat{q}_L$ by $\tau_n + 2 \ln n$. Plotting this correction as a function of $n$ and comparing to the expected variance of $\tau_n + 2 \ln n$ will indicate if the data is consistent with noise and thus whether the result of self-calibration is reliable. 

To perform self-calibration there is thus one parameter to be chosen, the index of the 
peak $n$. We will explore this choice in depth in Section \ref{sec:results}.
Meanwhile, it is instructive to consider Eq.~\ref{qap} for $n=2$. In this case, the self-calibrated $\hat{q}_S$
is simply given 
by the $\chi^2$ difference between the highest 
and second highest peaks (using $\tau_n \approx q_L^{(n)}$), apart from small
logarithmic corrections. If the highest peak is known to be physical, and the remaining peaks noise, this would give the $\chi^2$ difference between the physical peak and the highest noise-induced 
peak. So, in its simplest form, self-calibration 
corresponds to computing the look-elsewhere-corrected $p$-value from this difference in $\chi^2$. Furthermore, for sufficiently large $\hat{q}_S$, the look-elsewhere-corrected chi-squared, $S^2$, approximately equals this difference in $\chi^2$.

It is sometimes the case that the $\chi^2$ is only known up to a constant factor: for example in periodograms used for radial-velocity exoplanet searches, unknown jitter effects mean the measurement errors are only known up to a constant factor \citep[see e.g.][]{Baluev_2008}. In such cases, it is common practice to consider different normalizations of the $\chi^2$ to cancel out this factor, resulting in different analytical formulae for the $p$-value for each choice of normalization. These formulae can be complicated and have dependence on the number of data points, $N_d$. Instead, we can use self-calibration to analyse such cases in a general manner. If we only know the chi-squared up some constant factor, $k$, we have $q_L = k q_L'$ and $\tau_n = k \tau_n'$. We can thus self-calibrate $k$ by applying equation \ref{qap} twice using the $n^{\rm th}$ and $m^{\rm th}$ peaks, resulting in the following simultaneous equation:
\begin{align}
    \bar{\hat{q}}_S 
    &= k(\hat{q}'_L-\tau'_n)-2 \ln n - (M-2) \ln \frac{\hat{q}_L'}{\tau'_n} \label{qap_k} \\
    &= k(\hat{q}'_L-\tau'_m)-2 \ln m - (M-2) \ln \frac{\hat{q}_L'}{\tau'_m}.
\end{align}
Solving for $k$ gives the self-calibrated $k$ as
\begin{equation}
    \bar{k} = \frac{2 \ln \frac{n}{m} + (M-2) \ln \frac{\tau_m'}{\tau_n'}}{\tau_m'-\tau_n'}.
    \label{k}
\end{equation}
This can then be substituted into Eq.~\ref{qap_k} to compute the self-calibrated $\hat{q}_S$, and in turn the significance using Eq.~\ref{S_selfcalib}.

\section{Results}
\label{sec:results}

In this section we apply self-calibration of the $p$-value to various astrophysical examples. We start with a search for a Gaussian peak in white noise in subsection \ref{sec:wn_sc}. This is a 
very common example when searching for a single event in the data: it could for 
example 
correspond to detecting 
a single transient in an exoplanetary lightcurve (e.g.~a starspot). We then consider the Lomb-Scargle (LS) periodogram in subsection \ref{sec:periodogram}, where we allow for data with non-fixed spacing to investigate the effects of aliasing on self-calibration.
We then study a more complex (non-linear) planet transit model in subsection \ref{sec:kepler}, where we apply it to a Kepler exoplanet search. Finally, in subsection \ref{sec:planck} we consider an example from cosmology, namely searching for oscillatory features in the primordial power spectrum using data from Planck \citep{Planck13I}.

\subsection{Single Transient}
\label{sec:wn_sc}

\begin{figure}
    \includegraphics[width=\linewidth]{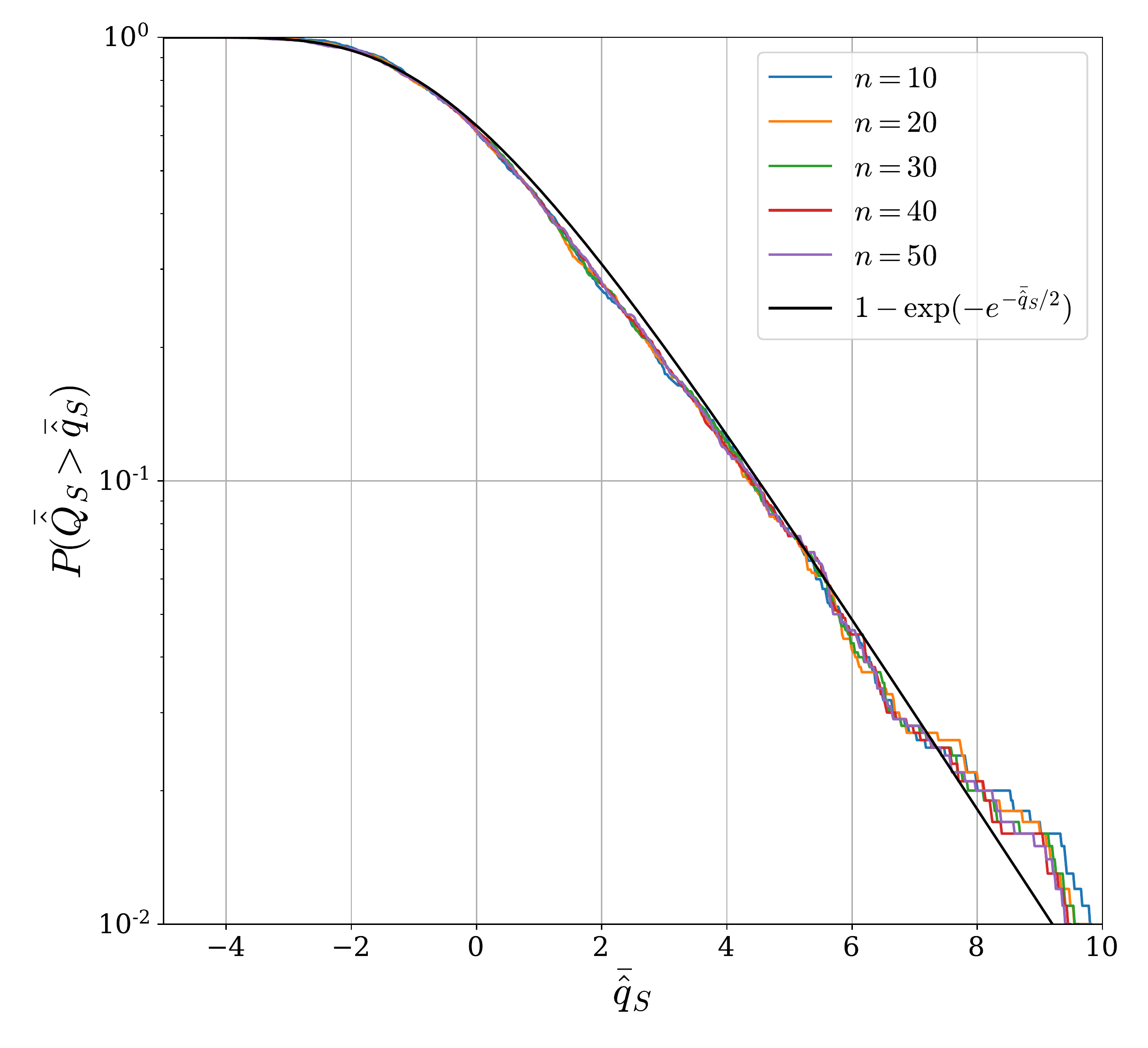}
\caption{
Complementary cumulative distribution of $\bar{\hat{q}}_S$ averaged over $10^3$ simulations with no signal ($A=0$). Self-calibration is performed with a variety of choices of $n$, all of which agree with the theoretical expectation of Eq.~\ref{sidak_selfcalib_qS} (black line). The noise at high $\bar{\hat{q}}_S$ is due to the finite number of simulations used.
}
\label{fig:selfcalib_pqa}
\end{figure}

\begin{figure*}
    \includegraphics[width=0.49\linewidth]{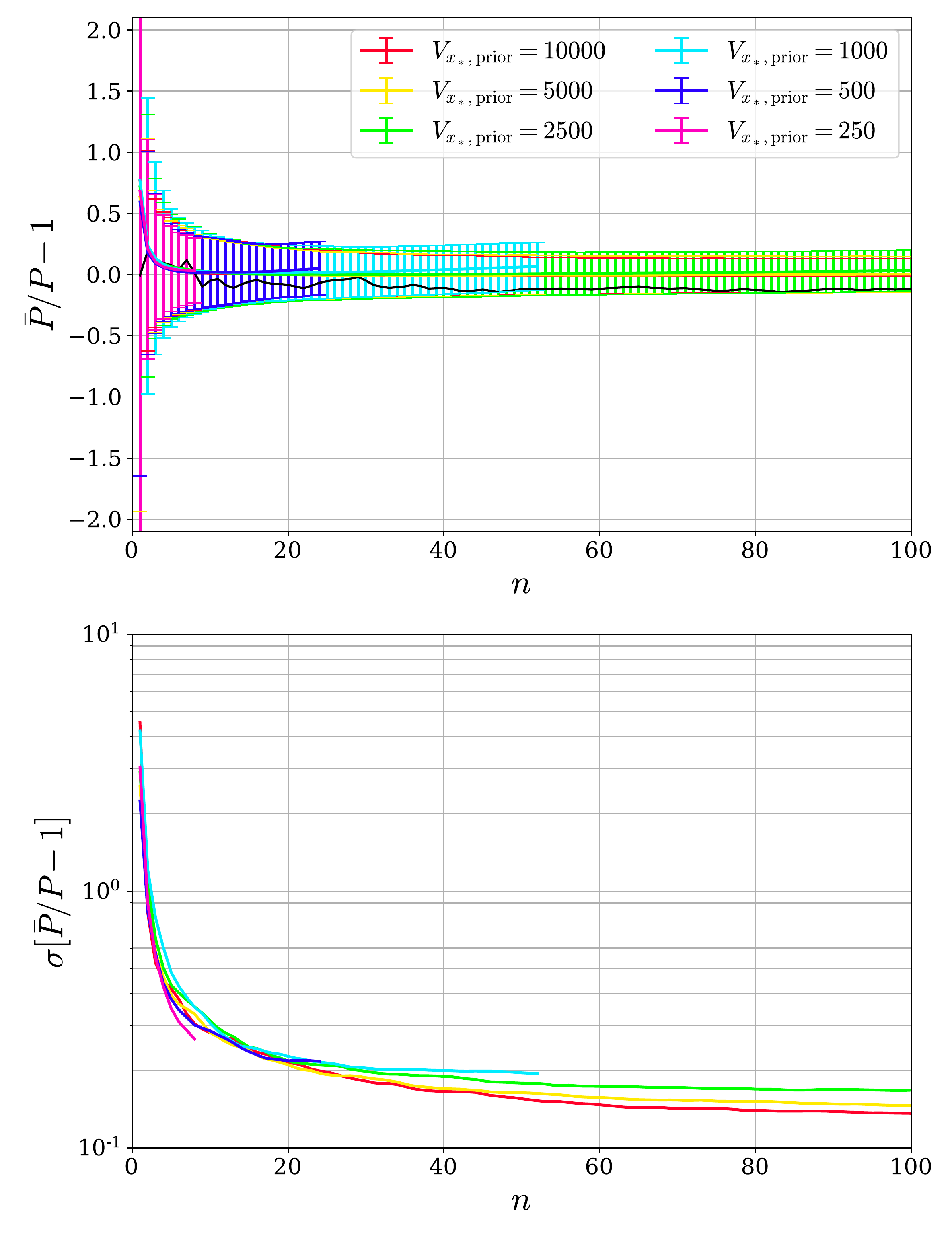}
    \includegraphics[width=0.49\linewidth]{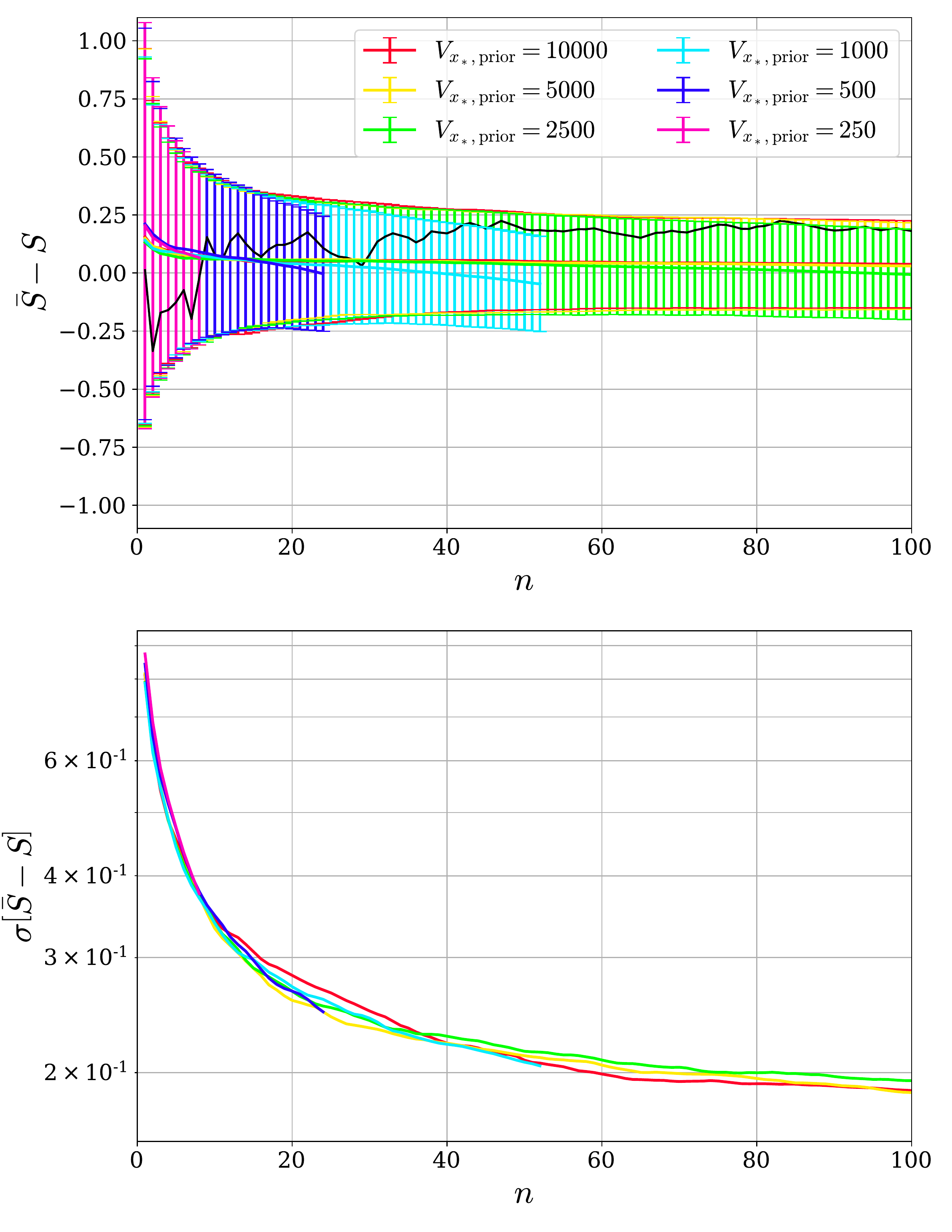}
    \caption{
    Analysis of the bias and variance of the self-calibrated $p$-value and significance as a function of $n$. 
    (Top left) The mean over $10^3$ realizations of $\bar{P} / {P} - 1$ as a function of $n$. Error bars represent the standard deviation. 
    (Bottom left) The standard deviation of $\bar{P} / {P} - 1$ as a function of $n$.
    Similarly,
    (top right) is the mean of $\bar{S}-S$ as a function of $n$, with 
    (bottom right) the standard deviation.
    In all cases 6 prior volumes are considered. 
    An example of a single realization for $V_{x_*, {\rm prior}} = 10^4$ is shown by the black line. 
    }
    \label{fig:selfcalib_delta_qasc_pqa_mu_std}
\end{figure*}

\begin{figure}
    \includegraphics[width=\linewidth]{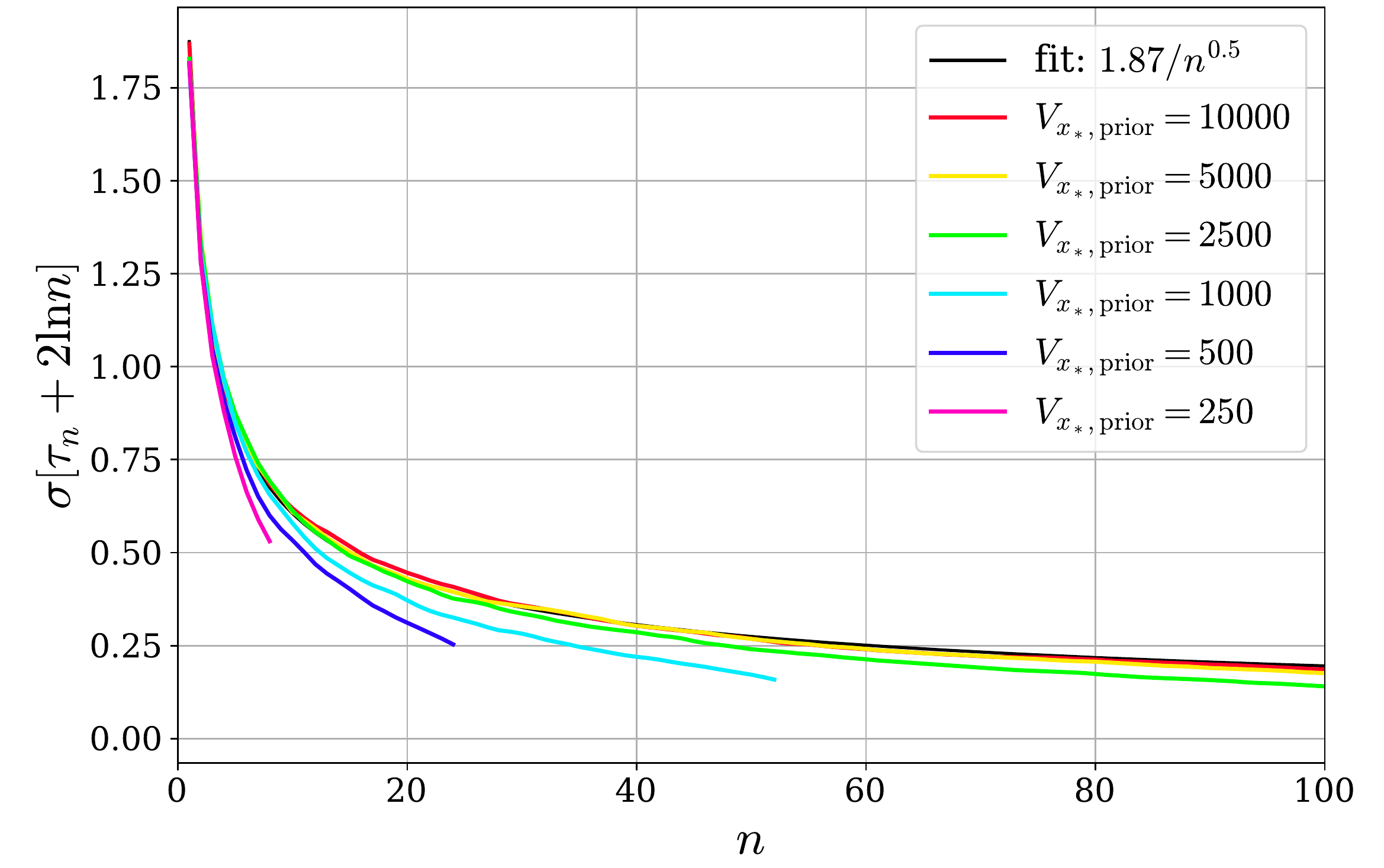}
\caption{
The standard deviation of $\tau_n + 2 \ln n$ as a function of $n$ for a variety of prior volumes. The black line shows a fit to the $V_{\rm prior}=10,000$ line given by Eq.~\ref{Qn_fit}.
}
\label{fig:selfcalib_Q_n_plus_std}
\end{figure}

We now apply self-calibration to an example of a search for a Gaussian peak in a white noise time series. Physically speaking, this could correspond to searching for Gaussian-like transients in lightcurves (e.g.~starspots), Gaussian-like peaks in spatial maps, as well as numerous other 
examples.
We consider a time series of data measurements $y(x)$ comprising of $N_d$ data points, $\bi{x}= \{x^i\}_{i=1}^{N_d}$, with spacing $x^{i+1} - x^i =1$. Each $y^i \equiv y(x^i)$ measurement has normally distributed noise
with zero mean and unit variance. We seek a signal of the form $A N(x | x_*, \sigma_*)$ where $A$ is the amplitude, and $N(x | x_*, \sigma_*)$ is a normal distribution with mean $x_*$ and width $\sigma_*$.
%
%
For these standard normal measurements, $q_L$ from Eq.~\ref{qL} equals the difference in chi-squared between the null and signal hypotheses:
\begin{align}
    q_L(\bi{x} | A, x_*, \sigma_*) = \sum_{i=1}^{N_d} ~ \left[ y^i \right]^2 - \left[ y^i - A N(x | x_*, \sigma_*) \right]^2.
\end{align}

To test self-calibration we first consider the case of pure noise.
To study different choices of $n$ we consider $10^3$ 
pure noise realizations. We use a signal width of $\sigma_* = 2$ such that the width of the peak is larger than the spacing of data points. Hence, this is a 2 dimensional problem, $M=2$, with parameters $A$ and $x_*$.
We consider 6 prior volumes in the range 250 to $10^4$. Note that $V_{x_*, {\rm prior}} = 10^4$ contains approximately $700$ peaks.

Fig.~\ref{fig:selfcalib_pqa} shows the complementary cumulative distribution of $\bar{\hat{q}}_S$ (Eq.~\ref{qap}) for different choices of $n$. There is excellent agreement with Eq.~\ref{sidak_selfcalib_qS} for all choices of $n$ for the full range of $p$-values, suggesting the self-calibrated $\hat{q}_S$ is distributed as the true $\hat{q}_S$.
To investigate the correspondence between the self-calibrated values and the values computed using the prior-to-posterior volume, we consider the bias and standard deviation of $\bar{P}/P-1$ and $\bar{S}-S$ on the left and right of Fig.~\ref{fig:selfcalib_delta_qasc_pqa_mu_std} respectively. It can be seen that the bias is close to 0, with variance decreasing with $n$ and roughly levelling off at around $n \sim 20-40$. The plots also show independence of prior volume, provided $n$ is sufficiently small. As $V_{\rm prior}$ is decreased, the number of peaks in the data decreases, and thus the maximum possible $n$ that can be used decreases. It can be seen that the smaller the prior volume, the earlier in $n$ self-calibration picks up a bias. This bias continues to grow with $n$, however we cut the lines for the sake of neatness. Thus the bias remains constant up to a particular value of $n$ which grows with $V_{x_*, {\rm prior}}$, i.e. the number of peaks in the data.

The variance of the $p$-value levels of at around $n=20$. At this $n$ the rms of the fractional error of the $p$-value is around 0.2, meaning the self-calibrated $p$-value will have an error of $\pm 20 \%$, corresponding to a error in $S$ of $\pm 0.25$.  We see that even for the case of $V_{x_*, {\rm prior}}=250$, which contains approximately $20$ peaks per realization, one can self-calibrate the $p$-value to $\pm 30 \%$ using $n=8$. 
This level of accuracy suffices for many applications where one wishes to quickly quantify the $p$-value.
A final feature of Fig.~\ref{fig:selfcalib_delta_qasc_pqa_mu_std} are the black lines in the upper plots, which show the bias for a single realization, where 
it can be seen that there is correlation between different values of $n$.

As discussed earlier in Section \ref{sec:selfcalib}, the rms of $\tau_n + 2 \ln n$ as a function of $n$ can be used as a diagnostic to verify whether the data is consistent with pure noise, and thus whether the result of self-calibration is reliable.
This is plotted in Fig.~\ref{fig:selfcalib_Q_n_plus_std}.  It can be seen that the variance is prior independent up to a value of $n$ that grows with $V_{x_*, {\rm prior}}$. The line for $V_{x_*, {\rm prior}}=10^4$ is suitably converged in the range $1 \leq n \leq 100$; a fitting formula for this line is given by 
\begin{equation}
    \sigma \left[ \tau_n + 2 \ln n \right] = \frac{1.87}{\sqrt{n}}.
    \label{Qn_fit}
\end{equation}
While this fitting formula was derived for the $M=2$ white noise model considered in this section, we expect it to still be a good indicator for other models.
Thus when performing self-calibration, we advise plotting $\tau_n + 2 \ln n$ against $n$ for the dataset, and comparing this with the error envelope obtained from the above fitting formula. If the line falls within the error envelope one can be confident that self-calibration was performed reliably. If not, it could indicate the presence of 
physical peaks in the 
data, in which case one must identify and remove these physical signals from the data, and then repeat the self-calibration process, to avoid an overly conservative estimate of the $p$-value.

\begin{figure*}
    \includegraphics[width=\linewidth]{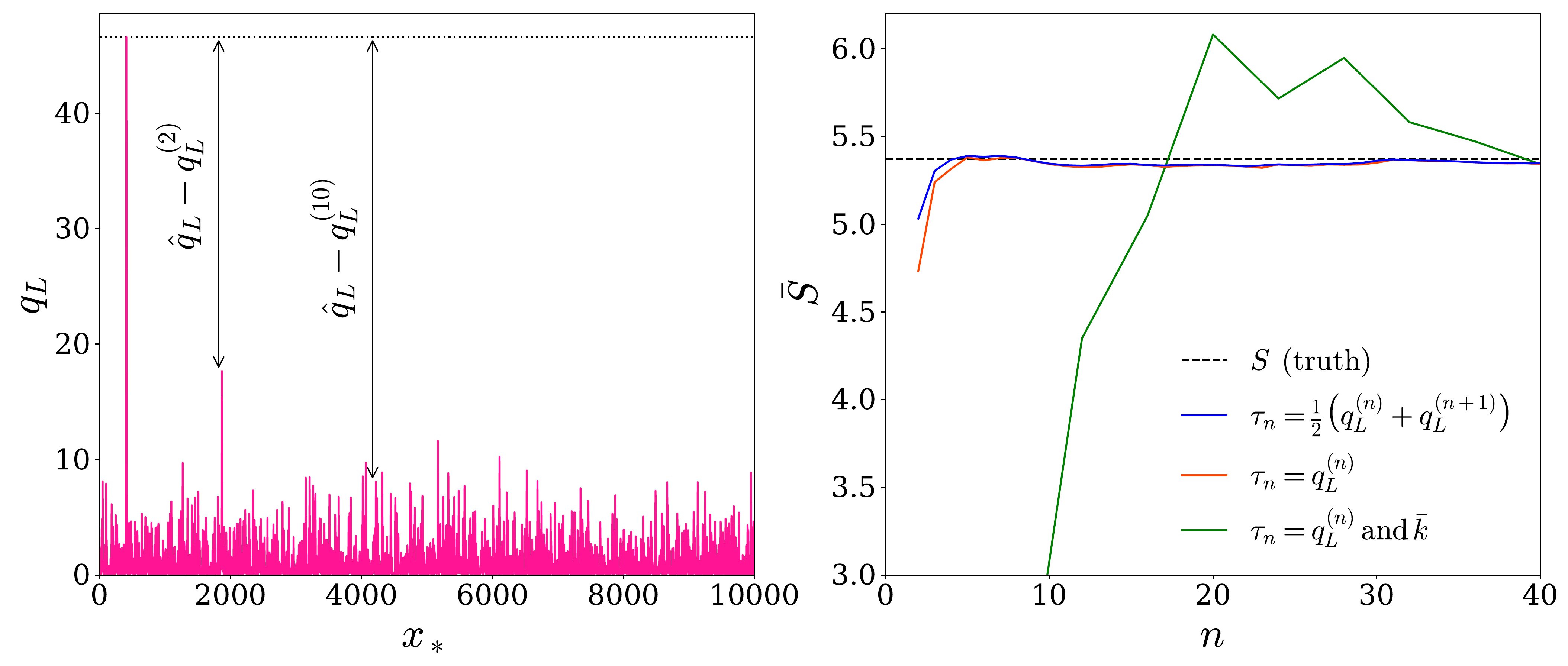}
\caption{
An example of self-calibration for a dataset with a true signal at $x_*=300$ and true significance $S=5.38$. \textit{Left panel}: Distribution of $q_L$ projected onto the $x_*$ axis, i.e. having maximized over $A$ for each $x_*$. The maximum $q_L$ is defined as $\hat{q}_L \equiv q_L^{(1)} \approx 47$, corresponding to a 6.8 sigma signal. Two arrows illustrate $\hat{q}_L - q_L^{(n)}$ for $n=2,10$, with $q_L^{(2)} \approx 18$ and $q_L^{(10)} \approx 9$. \textit{Right panel}: The self-calibrated value of the statistical significance $\bar{S}$ of the peak
to be a true peak, as a function of $n$ with two options for $\tau_n$ (blue and orange). We see these lines are well converged for $n\gtrsim5$. We also plot the results of self-calibration when one does not know the normalization of $q_L$ and must additionally self-calibrate its normalization, denoted by $k$ (green). In this case results are noisier --- due to the error in the estimate of $k$ --- but are still within 1-sigma for $n\gtrsim15$ and within 0.1-sigma for $n\gtrsim40$.
}
\label{fig:selfcalib_5sigma}
\end{figure*}

\begin{table*}
\begin{tabular}{|l|}
\hline
\textbf{SELF-CALIBRATING THE \textit{P}-VALUE} \\
\hline
 (1) \,
 By scanning over $\bi{z}_{>1}$ compute the likelihood ratio, and in turn $q_L$, of all high peaks. Denote the highest as $\hat{q}_L$.\\
 (2) \, 
 Choose $n$ (as discussed above). \\
 (3) \,
 Compute $\tau_n$, the average $q_L$ of the $n^{\rm th}$ and $(n+1)^{\rm th}$ highest noise peaks.\\
 (4) \,
 Evaluate the $p$-value of the highest peak as $1-\exp \left( -e^{-\bar{\hat{q}}_S/2} \right)$, where $\bar{\hat{q}}_S = \hat{q}_L-{\tau}_n-2 \ln n - (M-2) \ln \frac{\hat{q}_L}{\tau_n}$. \\
 (5) \,
 Evaluate the statistical significance, or number of sigma, approximately as $\bar{S}=\sqrt{\bar{\hat{q}}_S-\ln 2 \pi \bar{\hat{q}}_S + 2 \ln t}$. \\
 (6) \,
 Plot ${\tau}_n+2 \ln n$ vs $n$ together with the theoretical error envelope to verify the peaks are consistent with noise.\\
 \textcolor{white}{(6)} \,
 If not, remove additional physical peaks and repeat. One can then also repeat the process to evaluate the significance of each physical peak. \\
\hline
\end{tabular}
\caption{Algorithm for self-calibration of the $p$-value and statistical significance.}
\label{tab:sc-pval}
\end{table*}

Having studied self-calibration in the context of pure noise, we now illustrate the method applied to data with a physical signal. Fig.~\ref{fig:selfcalib_5sigma} shows a typical example of a peak at $x_*=300$ with true significance $S \approx 5.4$. The left plot shows the distribution of $q_L$ projected onto the $x_*$ dimension. It can be seen that there are numerous peaks, including a physically injected peak at $x_* \approx 300$ with $\hat{q}_L \equiv q_L^{(1)} = 46.6$. The right of Fig.~\ref{fig:selfcalib_5sigma} shows the self-calibrated significance compared to the true significance, as a function of $n$ and for different choices of threshold $\tau_n$ given in Eq.~\ref{tau}.
The simplest form of self-calibration is to use the difference in height of the highest two peaks, i.e. $\tau_n =q_L^{(n)}$ with $n=2$. Using Eq.~\ref{qap} this gives $\bar{\hat{q}}_S = \hat{q}_L - q_L^{(2)} - 2 \ln 2$. 
It can be seen that such an approach leads to under-predicting the number of 
sigma $S$ by $0.6$, so even at $n=2$ we 
obtain a useful diagnostic. Nevertheless, to achieve a better estimate of $S$ one can use higher $n$. By considering the blue line it can be seen that the average of the $1^{\rm st}$ and $2^{\rm nd}$ noise peaks produces a better estimate than the $1^{\rm st}$ noise peak alone, however taking such an average becomes unnecessary for larger $n$, and convergence is achieved by around $n=5$. This is typical for peaks of significance 5 sigma and above, thus for cases of physical interest one can often use very low values of $n$.

We also show the self-calibrated number of sigma for when the normalization of $q_L$ is unknown. While this is not a typical example, this is often the case when the measurement error is not known in radial-velocity exoplanet searches \citep[see e.g.][]{Baluev_2008}. To tackle such problem with self-calibration, one can self-calibrate using 2 peaks to estimate the normalization of $q_L$, $k$, using Eq.~\ref{k}. While one is free to use any choice of peaks, indexed by $n$ and $m$, we choose $m=n/3$ for this plot --- it is also typically good practice to choose $|m-n|\gtrsim5$ to avoid correlation between adjacent peak heights. The self-calibrated $\bar{S}$ in this case is depicted with the green line in Fig.~\ref{fig:selfcalib_5sigma}. It can be seen that there is now more noise in the estimated number of sigma, due to the noise introduced by having to estimate $k$, however results are still within 1-sigma of the correct value 
for $n\gtrsim15$ and within 0.1-sigma for $n\gtrsim40$. 

In summary, we have shown that one can accurately compute the look-elsewhere corrected $p$-value by considering the heights of likelihood peaks, without needing to evaluate the posterior volume or performing simulations. This is true even in the non-asymptotic limit.
One can reliably self-calibrate the $p$-value to order 10\% accuracy, or equivalently the number-of-sigma significance to $\pm 0.25$, using a single dataset. 
The self-calibration algorithm for the $p$-value
is summarized in Table \ref{tab:sc-pval}. 

\subsection{Lomb-Scargle Periodogram}
\label{sec:periodogram}

The Lomb-Scargle (LS) periodogram \citep{Lomb_1976, Scargle_1982} (see \cite{VanderPlas_2018} for a review) corresponds to a search for a sine wave in a white noise background. Thus instead of the Gaussian signal considered in the previous subsection, we now consider a signal of the form $A \sin(\omega t + \phi)$, where $t$ is time, $\omega=2\pi/f$ is the angular frequency of the orbit, and $\phi$ is the phase. For standard normal measurements, $y^i$, $q_L$ from Eq.~\ref{qL} equals the difference in chi-squared between the null and signal hypotheses:
\begin{align}
    q_L(\bi{t} | A, \omega, \phi) = \sum_{i=1}^{N_d} ~ \left[ y^i \right]^2 - \left[ y^i - A \sin(\omega t + \phi) \right]^2.
\end{align}
Physically this could correspond to the radial velocity of a star caused by a planet orbiting it \citep{Baluev_2008}, to the spectral analysis of solar neutrinos \citep{Ranucci_2007}, or to many other examples.

One of the key features of the LS periodogram over the classical periodogram is its consideration of non-uniform time measurements. For a time series with span $T$, we thus consider $N_d$ time measurements $\bi{t}= \{t^i\}_{i=1}^{N_d}$ between $0 \leq t < T$, with both fixed-uniform spacing and randomly distributed measurements. 
For data with fixed-uniform spacing, there is no information beyond the Nyquist frequency, $f_{\rm Nyq} = 1/(2\Delta t)$, where $\Delta t = (N_d-1)/T$ is the uniform data spacing. All frequency features beyond the Nyquist are indistinguishable from their aliases in the $f<f_{\rm Nyq}$ region, so 
in the case of fixed-uniformly space data, the aliasing follows a periodic pattern with period $2 f_{\rm Nyq}$.\footnote{Note the LS periodogram contains an additional symmetry $P(f)=P(-f)$, thus $P(f)=P(2f_{Nyq}-f)$, and there is no information beyond $f_{\rm Nyq}$.} 
Moreover, for 
uniform sampling we only need to 
consider the discrete frequencies
$f_i=i/(2 N_d \Delta t)$, and since Fourier modes
are orthogonal on this uniform 
basis the amplitudes of the different frequencies are independent of each other. In this situation we can 
fit for each frequency separately and 
our self-calibration can be applied  without error.

However, data without fixed-uniform spacing can extract frequency features beyond the Nyquist frequency.
As such, the periodic signal associated with a periodogram can introduce aliasing, whereby a physical signal at a particular frequency will give rise to peaks at both the true frequency and various other frequencies. Moreover, the peaks 
become correlated, and a proper
analysis requires a joint fit of all 
the peaks \citep{Foster_1995}. 
Instead, we can try to analyze the peaks 
individually without the joint analysis, ignoring the correlations: 
we will show that this still provides reasonably accurate results. We refer the reader to the work of \cite{Baluev_2008} for a detailed mathematical description of the effects of correlations/aliasing on the $p$-value in the context of the LS periodogram.

\begin{figure}
    \centering
    \includegraphics[width=\linewidth]{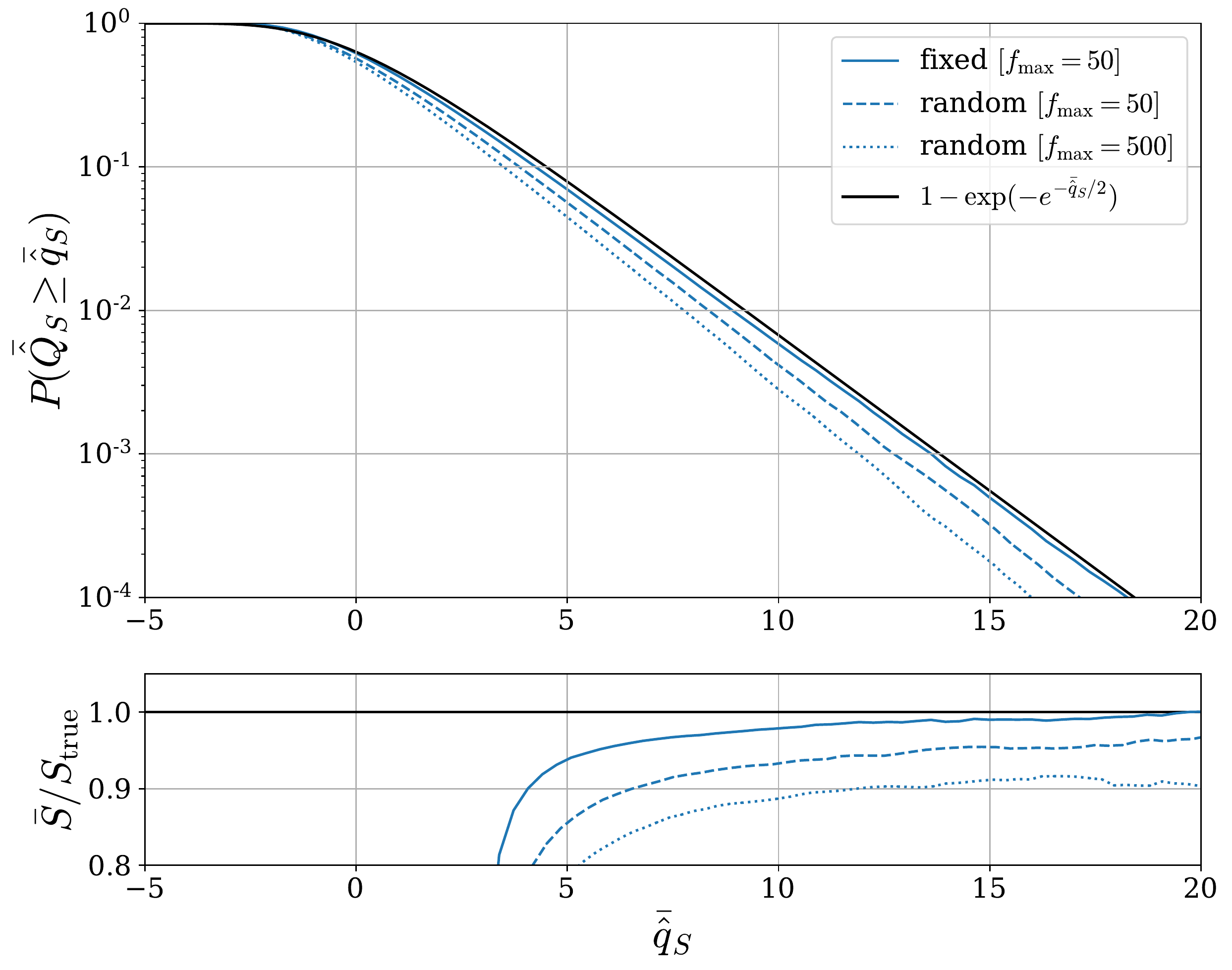}
    \caption{
    \textit{Top panel}: Distribution of self-calibrated $\hat{q}_S$ for the LS periodogram with $f_{\rm Nyq}=50$. We consider self-calibration using $n=10$, but have checked results are robust to this choice. We plot lines for fixed data spacing with $f_{\rm max}=f_{\rm Nyq}$ (solid), random data spacing with $f_{\rm max}=f_{\rm Nyq}$ (dashed), and random data spacing with $f_{\rm max}=10f_{\rm Nyq}$ (dotted).
    The fixed-spacing case agrees well with Eq.~\ref{sidak_selfcalib_qS}, while the agreement slightly deteriorates for randomly spaced data, and further as $f_{\rm max}$ is increased beyond the Nyquist. This discrepancy is introduced by aliasing effects, 
    however the \textit{bottom panel} shows the asymptotic fractional error on the number of sigma, $S$, is at most 10\%.
    }
    \label{fig:periodogram}
\end{figure}

We consider an LS periodogram with time of observation $T=1$, and $N_d=100$ measurements. The Nyquist frequency is thus $f_{\rm Nyq}=50$. We additionally apply a low frequency cutoff of $f_{\rm min}=0.5$ corresponding to the minimum frequency detectable for $T=1$. We use $10^6$ realizations to numerically compute the distribution of the self-calibrated $\hat{q}_S$ and in turn the $p$-value.

The left panel of Fig.~\ref{fig:periodogram} shows the distribution of self-calibrated $\hat{q}_S$ for this LS periodogram. We consider self-calibration using $n=10$, but have checked results are robust to this choice. The solid blue line corresponds to fixed-uniform data spacing with $f_{\rm max}=f_{\rm Nyq}$.
It can be seen that there is good agreement with Eq.~\ref{sidak_selfcalib_qS}: this is to be expected as the maximum of the LS periodogram is known to correspond to a chi-square with 2 degrees of freedom at any fixed phase \citep{Scargle_1982}, and so obeys Eq.~\ref{pql_global}. 

The dashed and dotted blue lines of Fig.~\ref{fig:periodogram} correspond to the $p$-value for random uniformly spaced data (i.e.~the data points $t^i$ are drawn from a uniform distribution). In this case we consider maximum frequencies of both 50 and 500, as non-uniformly space data produces information beyond the Nyquist frequency.
It can be seen that the results of self-calibration for $f_{\rm max} = f_{\rm Nyq} = 50$ shows slightly worse agreement than the fixed-spacing case, and that the agreement worsens as we increase to $f_{\rm max} = 10 f_{\rm Nyq} = 500$.
The reason for this worsening is that considering non-uniformly spaced data, and frequencies above the Nyquist, introduces additional peaks to the likelihood, which are correlated with each other. 
For this example the self-calibrated number 
    of sigma $S$
is still correct to within $10\%$ in the asymptotic limit, 
as shown in the bottom panel of Fig.~\ref{fig:periodogram}. 
Furthermore, since self-calibration underestimates the statistical significance it could lead to false negatives, but not false positives, so it is a conservative 
estimate. Given the difference between the local and global $p$-value is often many orders of magnitude, self-calibration assuming independent peaks still provides a useful fast way to approximate the $p$-value in cases of data without fixed-uniform spacing.

To illustrate why a joint analysis would be needed, consider fixed-uniform data spacing, for which the likelihood peaks in the range $(0,f_{\rm Nyq}]$ will repeat themselves in each  $(i f_{\rm Nyq}, (i+1)f_{\rm Nyq}], \, i\in  \mathbb{Z} $ region due to aliasing. This means that if one were to consider $f_{\rm max}>f_{\rm Nyq}$ there would be multiple repeated peaks of the same height and using  self-calibration on individual peaks assuming they are uncorrelated would break down. One can instead do a 
joint fit of all the peaks. In this 
specific example this is equivalent 
to removing the signal associated with the maximum peak from the data before computing $q_L^{(2)}$, and then iteratively removing the signal of each peak to get to higher $n$. This will remove all the peaks caused by aliasing, and generalizes self-calibration to any frequency range in the case of fixed-uniform data spacing. While this example is not of physical interest, as there is no extra information beyond the Nyquist frequency for fixed-uniform data spacing, it motivates the solution to aliasing in cases of non-uniform data spacing. In our experiments 
we found this procedure of removing the peaks tends to remove too much signal from the higher 
order peaks when the peaks 
are correlated, so that 
self-calibration is not very 
accurate. For this reason 
we argue the correct procedure 
is to 
fit for all the peaks jointly, 
which is computationally expensive and beyond the scope of this work. 

\subsection{Kepler Exoplanet Search}
\label{sec:kepler}

\begin{figure}
    \centering
    \includegraphics[width=.84\linewidth]{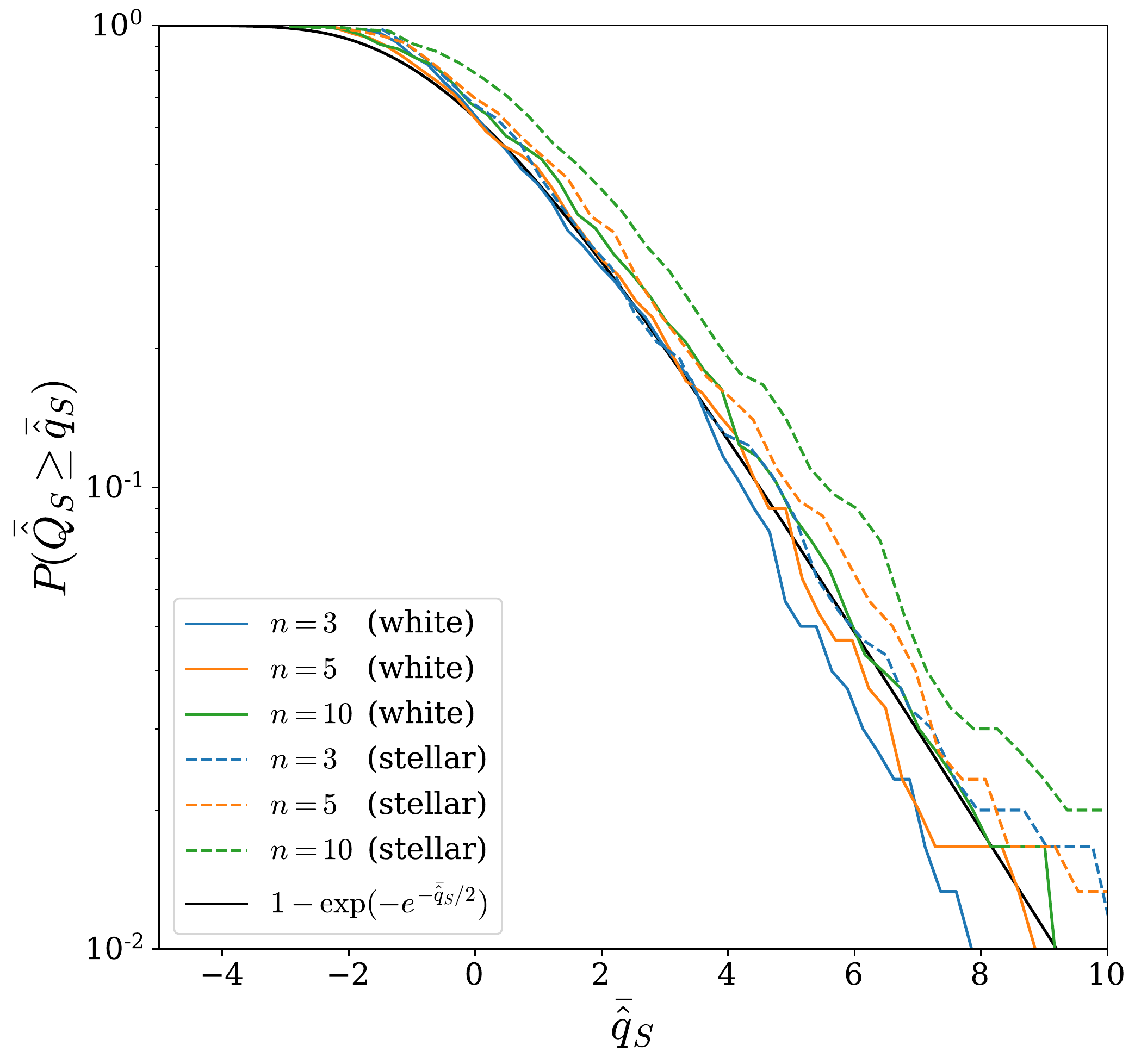}
    \caption{
    Distribution of self-calibrated $\hat{q}_S$ for exoplanet searches in Kepler-like data. We consider both white noise (solid) and realistic Kepler 90 stellar variability (dashed). We self-calibrate using $n=3,5,10$, and compare with Eq.~\ref{sidak_selfcalib_qS}.}
    \label{fig:keplerpvalue}
\end{figure}

Here we apply self-calibration in the context of exoplanet searches in the Kepler Space Telescope data \citep{KeplerSpaceTelescope}. The look-elsewhere effect is particularly prominent in these applications: in a pure-noise simulation a typical highest noise peak will have signal-to-noise ratio of $5.5$, i.e. $q_L\approx30$, corresponding to 
multiplicity, or trials factor, in excess of 
$10^7$.

We have some time series data of fluxes $\{ y_i \}_{i = 0}^{N_d}$ , measured at discrete time points $t_i$, which are evenly spaced $t_i = i \Delta$. The flux is composed of a signal and noise,
\begin{equation}
    y_i = s(t_i) + \mathcal{N}_i.
\end{equation}
We consider two noise scenarios: (i) when $\mathcal{N}_i$ is assumed to be a normally distributed random variable with zero mean and unit variance which is not correlated with other data points (i.e.~white noise), and (ii) adding realistic Kepler 90 stellar variability.
Here we consider a signal $s(t)$ comprised of a periodic train of $\mathcal{T}$ transits with period $P$. The signal has $M = 3$ parameters, $z = (A, P, \phi)$: amplitude, period and phase, respectively. The form of the signal is given by
\begin{equation} \label{eq:signal}
    s(t \vert A, P, \phi) = A \sum_{r = 1}^{\mathcal{T}} U\bigg( \frac{t - (r+\phi) P}{\tau_K (P)}\bigg),
\end{equation}
where $U(x)$ is a U-shaped transit template which is nonzero in the region (-1/2, 1/2). We use Kepler's third law to give the duration of each transit event as $\tau_K \propto P^{1/3}$, effectively assuming that planet's orbits are circular and perfectly aligned with the line of sight. 

Following the analysis of \cite{GMF,robnik2019kepler}, matched filtering the data $y$ with the template $s_0(t \vert z)$ gives the signal-to-noise, $\sqrt{q_L}$, as
\begin{equation}
    q_L(\phi \vert P)^{1/2} = \mathcal{F}^{-1} \bigg\{ \frac{\FT{y}^* \ \FT{s_0}}{\mathcal{P}} \bigg\},
    \label{mf}
\end{equation}
\begin{equation}
    {\rm SNR} = \mathcal{F}^{-1} \bigg\{ \frac{\FT{d}^* \ \FT{s}}{\mathcal{P}} \bigg\},
\end{equation}
where $\FT{\cdot}$ is the discrete Fourier transform and $\mathcal{P}$ is the noise power spectrum (which for white noise is a constant equal to the number of data points $N_d$).
The template is normalized such that $\sum_{i = 0}^{N_d} |\FT{s_0}_i|^2 / \mathcal{P}_i = 1$.

We consider a star like Kepler 90, where data spanning 1465.6 days of observations with $\Delta = 29.4$ minute intervals is available. To test self-calibration for such a model, we consider 300 noise-only simulations.
We simulate time series, apply matched filtering, and search over periods in the range 3--300 days and over all phases.

Fig.~\ref{fig:keplerpvalue} shows the distribution of the self-calibrated $\hat{q}_S$ over these noise realizations for a few choices of peak index, $n$. The solid lines correspond to the case of white noise, for which it can be seen that there is good agreement with Eq.~\ref{sidak_selfcalib_qS}, and thus self-calibration produces accurate results for all $n$ considered.

Next we add realistic Kepler 90 stellar variability to the model, as shown by the dashed lines of Fig.~\ref{fig:keplerpvalue}. In this case the model decides whether to fit for the stellar variability or the exoplanet, or both. The null hypothesis is now 
noise and stellar variability, on 
top of which we are looking for 
signatures of exoplanets. We 
model stellar variability as a 
Gaussian process, measuring 
first its power spectrum from the 
data directly \citep{robnik2019kepler}, 
and then fitting for all of the 
Fourier components of the stellar 
variability (approximately 70,000 components). It has been shown in 
\cite{GMF} that the results of 
this joint fit are equivalent to 
the matched filter analysis, where 
we use the power spectrum for $\mathcal{P}$ in Eq.~\ref{mf}. We assume that the different 
planet peaks do not interact with each other. 
This makes such analysis feasible, 
unlike for the periodogram case where 
a joint fit of multiple peaks would 
be very expensive. 
Fig.~\ref{fig:keplerpvalue} shows
good agreement between self-calibration and simulations in the case of stellar variability, 
although there is a slight 
discrepancy for large $n$, so for an optimal 
analysis it suffices to use $n=5$. For 
Kepler data we can scramble the data mixing up different time 
intervals, which destroys exoplanet periodicity, 
and guarantees that we have pure noise peaks 
in the scrambled data, so we do not need to 
worry about presence of real planets in the 
lower amplitude peaks. 

\subsection{
Searching for oscillatory features in the primordial power spectrum}
\label{sec:planck}

\begin{figure*}
    \includegraphics[width=\linewidth]{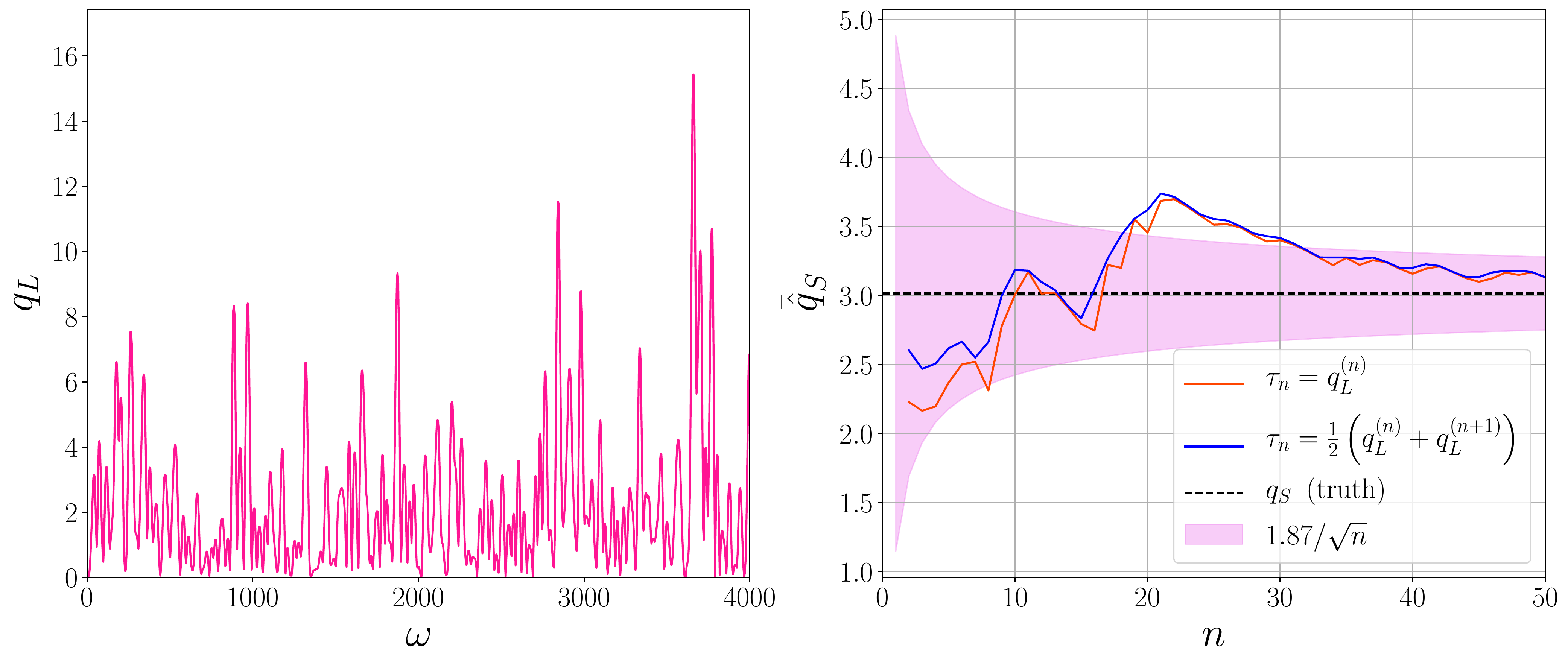}
\caption{
Example of self-calibration when searching the primordial power spectrum for oscillatory features. \textit{Left panel}: Distribution of $q_L$ projected onto the $\omega$ axis, i.e. having maximized over $A$ and $\phi$ for each $\omega$. The highest peak is at $\omega \approx 3660$, with $\hat{q}_L=15.4$, giving an uncorrected significance of $\sqrt{15.4} \approx 4$ sigma. \textit{Right panel}: The self-calibrated value of $\hat{q}_S$ for the 
highest peak as a function of $n$, with two options for $\tau_n$ from Eq.~\ref{tau} and theoretical error envelope from Eq.~\ref{Qn_fit}. The true value of $\hat{q}_S$ from using the posterior volume is $\hat{q}_S=3.0$, and self-calibration shows good agreement for all $n$. Using Eq.~\ref{sidak_selfcalib_qS}
with $\hat{q}_S=3$ gives the $p$-value as 0.20, giving a significance of $S=1.3$; this suggests that uncorrected 4-sigma 
peaks, such as this one, arise relatively commonly 
from noise fluctuations.}
\label{fig:selfcalib_planck}
\end{figure*}

Here we apply self-calibration to real data, in the context of a search for oscillatory features in the  primordial power spectrum. This is an example of a non-Gaussian model of cosmological inflation considered by \cite{Fergusson:2014hya, Fergusson:2014tza}. The model considered adds an oscillatory perturbation to the $\Lambda$CDM power spectrum as follows,
\begin{equation}
    P(k) = P_0(k) [1 + A \sin(2 \omega k + \phi)],
    \label{Pfeature}
\end{equation}
where $P_0(k)$ is the featureless ($\Lambda$CDM) power spectrum and $A$, $\omega$, and $\phi$ are the amplitude, frequency, and phase of the oscillatory perturbation. This is thus an example of an $M=3$ model. In this case one is uncertain what the frequency or phase of the oscillation is, and one scans over a large range of frequencies to seek a fit to theory -- introducing a large look-elsewhere effect.
Using the Planck 2013 likelihood \citep{Planck13_L}, we follow the analysis detailed in §6 of \cite{LEE1} to produce the likelihood distribution on the left of Fig.~\ref{fig:selfcalib_planck}. It can be seen to be highly multimodal with many spurious peaks. The maximum occurs as $\omega \approx 3660$ with $\hat{q}_L=15.4$. 
Note there are three additional peaks with $\hat{q}_L \geq 10$, and 
several more between 8 and 10. 

The true value of $\hat{q}_S$, obtained using the posterior volume, was found to be $\hat{q}_S = 3$. The right of Fig.~\ref{fig:selfcalib_planck} illustrates self-calibration, as detailed in Table \ref{tab:sc-pval}, showing that a sufficiently accurate approximation of $\hat{q}_S$ is achieved for all choices of $n$. Furthermore, the pink shaded region represents the fitting formula for the error envelope presented in Eq.~\ref{Qn_fit}, and it can be seen that it encloses the data well. This suggests that self-calibration is reliable and in general this can be used as a diagnostic, as one typically will not have the true value of $\hat{q}_S$ to compare with. It is useful to note that even though Eq.~\ref{Qn_fit} was obtained in the context of an $M=2$ white noise example, it still provides a useful diagnostic for different models.

\section{Conclusions}
\label{sec:conclusions}

This paper presents a new
method, \textit{self-calibration}, 
to compute statistical significance in the presence of the look-elsewhere effect by considering only the heights of peaks 
in the data likelihood distribution. 
These peaks are a byproduct of any peak-hunting
data analysis, so there is negligible computational cost in this approach. In contrast, 
existing methods rely on simulations, model-dependent analytical calculations, or explicit evaluation of the Bayes factor, all of
which can be 
time consuming.

In its simplest form, self-calibration subtracts the $\chi^2$ of the highest noise peak, typically 
assumed to be the second peak, 
from the $\chi^2$ of the highest peak, to approximate the look-elsewhere-corrected $\chi^2$ of the highest peak. Accuracy is 
improved by considering lower peaks, typically at a negligible computational cost since they are also a 
byproduct of the analysis. This approach assumes that these lower peaks are dominated by noise; when this is not the case, i.e.~there are multiple physical peaks, 
one can iteratively remove physical peaks from the data after 
verifying that their $p$-value 
is small. One can then also self-calibrate the significance of each physical peak. This is another reason to favor low 
amplitude peaks for the subtraction: one may not 
be certain if the highest peaks are physical or not, 
but one is often certain that the lower peaks 
are generated by noise. 
An alternative approach to noise-only 
peaks is to 
use some form of scrambled data where 
we know the signal has been eliminated. 
For example, in the exoplanet detections from transits this could 
either be an inverted or 
scrambled time series which 
eliminates the periodicity of the 
planet transits.

We showed that self-calibration gives an accurate estimate
of the 
FAP, or
$p$-value, of the 
highest peak(s) in various astrophysical examples, including planet detection, periodograms, and cosmology. We also 
developed a version of self-calibration which can be applied when the 
noise and likelihood are not 
known, where one must also 
self-calibrate the normalization. 
Our approach is general, but 
approximate: 
there are situations where fitting for individual peaks is inaccurate, and a joint fit accounting for the effects of correlated peaks is required. An example is periodogram analysis in the case of non-uniform data spacing. This  is not conceptually any different in the sense that if peaks are correlated then a joint fit is required, but 
it is computationally difficult, and 
for specific situations such as 
periodograms methods have been developed 
where one can account for these 
effects without doing a joint fit \citep{Baluev_2008}.
However, we have demonstrated that even without correcting for these effects, self-calibration provides a simple method to quickly determine whether a significant detection has been made, and is thus complementary
to the more specialized methods 
that apply to specific
situations.

\section*{Acknowledgements}
We are grateful to the anonymous referee for providing many useful comments that helped cultivate the paper into its final form.
AEB thanks Omer Ronen for insightful discussion on the manuscript.
This research made use of the Cori supercomputer at the National Energy Research Scientific Computing Center (NERSC), a U.S.~Department of Energy Office of Science User Facility operated under Contract No.~DE-AC02-05CH11231.
We acknowledge the use of CAMB \citep{Lewis_2000} and CosmoMC \citep{Lewis_2002} for the analysis in Section \ref{sec:planck}. 
This material is based upon work supported by the National Science Foundation under Grant Numbers 1814370 and NSF 1839217, and by NASA under Grant Number 80NSSC18K1274.

\section*{Data Availability}
The data products associated with the Kepler data analysis in Section \ref{sec:kepler} are available in the NASA Exoplanet Archive, at  \url{https://exoplanetarchive.ipac.caltech.edu/bulk_data_download/}.
The data products pertaining to the Planck 2013 analysis in Section \ref{sec:planck} are documented in \cite{Planck13I}.
Other data associated with this article will be shared upon reasonable request to the corresponding author.



\bibliographystyle{mnras}
\bibliography{refs}





\appendix

\section{Bayesian derivation of self-calibration}
\label{app:selfcalib}

Here we provide a Bayesian derivation of the self-calibration equations presented in Section \ref{sec:selfcalib}, based on the work of \cite{LEE1}.

From a Bayesian perspective one considers the Bayes factor, which considers the evidence ratio for the hypothesis that there is a signal $H$ to the null hypothesis $H_0$. 
The Bayes factor is the ratio of Bayesian evidence under each hypothesis:
\begin{equation}
    p(\bi{x}|H)=\int d\bi{z} ~ p(\bi{z}|H)p(\bi{x}|\bi{z},H), 
    \label{BFgen1}
\end{equation}
where $p(\bi{z}|H)$ is the prior under hypothesis $H$ and $p(\bi{x}|\bi{z},H)$ is the likelihood under hypothesis $H$.
Thus the Bayesian evidence is equal to the prior-weighted average of the likelihood.
For the examples considered in this paper we consider a null hypothesis for which the parameters are fixed, thus the evidence for $H_0$ is simply given by the null likelihood $p(\bi{x}|\bi{z}_{0})$.

For a multimodal likelihood, the Bayes factor can be approximated 
by performing a local integration at each peak. This leads to a sum over contributions from each posterior mode. If the location of the $\ell^{\rm th}$ highest mode is $\bi{z}=\bi{\mu}^\ell$, this gives Bayes factor as
\begin{equation}
    B
    \equiv \frac{p(\bi{x}|H)}{p(\bi{x}|H_0)}
    \approx \sum_\ell b^\ell,
    \label{BFgen}
\end{equation}
where each $b^\ell$ is the contribution of mode $\ell$ to the Bayes factor 
and can be parameterized as
\begin{equation}
    b^\ell
    = \frac{p(\bi{x}|\bi{\mu}^\ell)}{ p(\bi{x}|\bi{z}_0) } \frac{ V_{\rm posterior}^\ell  }{ V_{\rm prior}^\ell  }
    = e^{q_L^\ell/2} \frac{ V_{\rm posterior}^\ell  }{ V_{\rm prior}^\ell  },
    \label{bl}
\end{equation}
where $V_{\rm prior}^\ell$ ($V_{\rm posterior}^\ell$) is the prior (posterior) volume associate with mode $\ell$.
A common approximation for these volumes is the Laplace approximation \citep{Laplace_approx}, in which case the prior volume is given by
\begin{equation}
    V_{\rm prior}^\ell \simeq 1/p(\bi{\mu}^\ell),
    \label{Vprior}
\end{equation}
and the posterior volume is given by 
\begin{equation}
    V_{\rm posterior}^\ell \simeq (2 \pi)^{M/2} \sqrt{\det \bi{\Sigma}^\ell},
    \label{Vpost}
\end{equation}
where $\bi{\Sigma}$ is the covariance matrix.
We note that in principle one can compute Bayes factor exactly, and the Laplace approximation is only a simple, but often effective, approximation \citep{LEE1}.

One can combine the frequentist and Bayesian perspectives to define a new test statistic
\begin{align}
    q_S \equiv q_L - 2 \ln N + \ln 2 \pi q_L - 2 \ln t,
    \label{qa}
\end{align}
where the trials factor is taken as the prior-to-posterior volume ratio for the parameters $\bi{z}_{>1}$ at the 
maximum peak,
\begin{equation}
N = \frac{\hat{V}_{\rm >1,prior}}{\hat{V}_{\rm >1,posterior}},
\label{N}
\end{equation}
and $t=1,2$ for a one, two-tailed test. 
Note that while this is the prior-to-posterior volume ratio for $\bi{z}_{>1}$, we will often simply refer to it as the prior-to-posterior volume.
The $p$-value is then given by 
\begin{equation}
    P(\hat{Q}_S > \hat{q}_S) \simeq 1-\exp \left (-e^{-\hat{q}_S/2} \right),
    \label{sidak}
\end{equation}
and applies both asymptotically and non-asymptotically. The key difference between the $p$-value expressions of equations \ref{pql_global} and \ref{sidak}  is that the latter has no explicit $N$ dependence, meaning the $p$-value in terms of $\hat{q}_S$ is unaffected by the look-elsewhere effect.
This makes $\hat{q}_S$ a more useful statistic to use.
The statistical significance, or the number of sigma, can be approximated as 
\begin{equation}
    S \approx \sqrt{\hat{q}_S - \ln 2 \pi \hat{q}_S + 2 \ln t},
    \label{S}
\end{equation}
with corrections of order $ \mathcal{O} (\hat{q}_S^{-3/2})$. The look-elsewhere-corrected chi-squared is $S^2$. Note for sufficiently large $\hat{q}_S$, $S \approx \sqrt{\hat{q}_S}$.

Thus all one needs to evaluate the 
$p$-value is $\hat{q}_S$, which itself depends on $\hat{q}_L$ and $N$. Computing $N$ requires the evaluation of the 
posterior volume over $\bi{z}_{>1}$, which can be evaluated using the Laplace approximation, Variational Inference or Monte Carlo Markov Chain methods. However, we seek a faster alternative.

The asymptotic scaling of the $p$-value with $N$, and thus the prior volume, in Eq.~\ref{sidak} 
offers a 
way to evaluate the trials factor from the distribution of $q_L$ across the peaks in a dataset. One can evaluate the 
$p$-value in subvolumes of the data by counting the number of peaks above some threshold, and then rescale this to give the $p$-value for the entire volume. 
Calibrating $N$ in this way is cheaper than evaluating the posterior 
volume directly, for example by using Monte Carlo methods, because peaks in the likelihood are a byproduct of the peak-search analysis. 
Moreover, provided the 
$q_L$ peaks are dominated by noise, one can perform this calibration on the 
data directly without needing to run simulations. We thus call this method self-calibration, as one is calibrating $N$ using the peaks belonging to the measured data itself.

We start by splitting the prior volume of $\bi{z}_{>1}$ into $K$ bins, such that the prior volume of one of the bins is 
\begin{equation}
    V_{\rm >1, prior}' = V_{\rm >1,prior} / K.
    \label{V_pri_bin}
\end{equation}
We note that for non-uniform priors, one may need to include correction terms in this splitting of the prior volume to obtain greater accuracy (see e.g.~\cite{Baluev_2015} for discussion of splitting complex prior volumes).

Thinking of the $p$-value in terms of the false-positive rate (FPR), one can approximate the $p$-value of a bin as the fraction of bins containing a peak with $q_L > \tau$, for some threshold $\tau$.
 Smaller peaks typically have a larger error on $\bi{z}_{>1}$, which scales approximately as the noise-to-signal, i.e. $\hat{q}_L^{-(M-1)/2}$. Thus the average posterior volume in one of the bins is related to the posterior volume of the full volume by 
 \begin{equation}
    V_{\rm >1,posteriror}' \simeq (\hat{q}_L/\tau)^{(M-1)/2} V_{\rm >1,posteriror}.
    \label{V_post_bin}
\end{equation} 

Substituting equations \ref{V_pri_bin} and \ref{V_post_bin} into Eq.~\ref{N} gives the trials factor for a single bin as
\begin{equation}
    N' \simeq \frac{N}{K} \left( \frac{\tau} {\hat{q}_L} \right)^{(M-1)/2}.
\end{equation}
%
Denoting the number of bins containing at least one peak with $q_L > \tau$ as $n_{\rm bins}(\tau)$,
the FPR is given by
the fraction of bins 
satisfying this condition. Equating the FPR to the $p$-value from Eq.~\ref{sidak} gives
\begin{align}
  &\frac{n_{\rm bins}(\tau)}{K}
  = 1-\exp \left(-e^{-\frac{1}{2}\left[\tau-2\ln N' + \ln 2\pi \tau - 2 \ln t \right]} \right) \\
  &{\rm \quad\,\,\,\,\,}= 1-\exp \left(-e^{-\frac{1}{2}\left[\tau-2\ln N + 2\ln K + \ln 2\pi \hat{q}_L + (M-2) \ln \frac{\hat{q}_L}{\tau} - 2 \ln t\right]} \right).
  \label{sc}
\end{align}
Rearranging and taking the $K \rightarrow \infty$ limit gives an expression for the trials factor
\begin{align}
  2 \ln N
  \rightarrow \tau + 2 \ln n_{\rm peaks}(\tau) + \ln 2 \pi \hat{q}_L + (M-2) \ln \frac{\hat{q}_L}{\tau} - 2 \ln t,
  \label{Nself_lim}
\end{align}
where 
the number of bins with a $q_L$ peak larger than $\tau$ tends to the number of peaks with $q_L$ larger than $\tau$ in the full volume, i.e. $n_{\rm bins}(\tau) \rightarrow n_{\rm peaks}(\tau)$ as $K \rightarrow \infty$.
This provides an estimate of $N$ that relies purely on the number and height of the peaks, without needing to evaluate their posterior volumes.\footnote{Note we have ignored boundary effects (see e.g.~\cite{Baluev_2013}), however these will be negligible in the case of large $N$, i.e.~when the look-elsewhere effect is considerable.}
Combining with Eq.~\ref{qa} this gives $\hat{q}_S$ as 
\begin{equation}
    \hat{q}_S=\hat{q}_L-\tau-2 \ln n_{\rm peaks}(\tau) - (M-2) \ln \frac{\hat{q}_L}{\tau},
    \label{qap_continuous}
\end{equation}
and Eq.~\ref{sidak} gives the $p$-value as
\begin{equation}
P(\hat{Q}_L>\hat{q}_L) =1-\exp \left(-e^{ -\frac{1}{2} \left[\hat{q}_L-\tau-2 \ln n_{\rm peaks}(\tau) - (M-2) \ln \frac{\hat{q}_L}{\tau} \right] } \right).
 \label{sidak_selfcalib_continuous}
\end{equation}

In the above formulae, $\tau$ is a continuous variable, however a single dataset consists of a discrete set of peaks. To self-calibrate the $p$-value of a particular dataset we thus evaluate the above formulae for the $n^{\rm th}$ peak such that
\begin{align}
    n_{\rm peaks}(\tau_n) &= n,
    \label{n}
\end{align}
with
\begin{align}
    \tau &= \tau_n \equiv \frac{q_L^{(n)} + q_L^{(n+1)}}{2} \approx q_L^{(n)},
    \label{tau}
\end{align}
where
$q_L^{(n)}$ is the $q_L$ value of the $n^{\rm th}$ highest
peak. 
We choose $\tau$ as the average of $q_L^{(n)}$ and $q_L^{(n+1)}$ as a simple way to account for the discreteness of the data: for example, if $q_L^{(n)} = 10$ and $q_L^{(n+1)} = 6$, one can only conclude that $n$ corresponds to $\tau$ in the range $6 < \tau \leq 10$. Having said that, in many cases one can choose $\tau_n = q_L^{(n)}$ and achieve sufficient accuracy. We explore both options of $\tau_n$ in the main paper.

Substituting equations \ref{n} and \ref{tau} into equations \ref{Nself_lim}, \ref{qap_continuous}, and \ref{sidak_selfcalib_continuous}, gives the self-calibrated estimate of $2 \ln N$ as
\begin{align}
  2 \ln \bar{N}
   \equiv \tau_n + 2 \ln n + \ln 2 \pi \hat{q}_L + (M-2) \ln \frac{\hat{q}_L}{\tau_n} - 2 \ln t,
  \label{tlnNsc_app}
\end{align}
the self-calibrated estimate of $\hat{q}_S$ as 
\begin{equation}
    \bar{\hat{q}}_S \equiv \hat{q}_L-\tau_n-2 \ln n - (M-2) \ln \frac{\hat{q}_L}{\tau_n},
    \label{qap_app}
\end{equation}
and the self-calibrated estimate of the $p$-value as
\begin{equation}
\bar{P}(\hat{Q}_L>\hat{q}_L) \equiv 1-\exp \left(-e^{ -\frac{1}{2} \left[\hat{q}_L-\tau_n-2 \ln n - (M-2) \ln \frac{\hat{q}_L}{\tau_n} \right] } \right),
 \label{sidak_selfcalib_app}
\end{equation}
where bars are used to indicate these expressions are the self-calibrated approximations.
Finally, Eq.~\ref{S} can be used to compute the self-calibrated significance as
\begin{align}
    \bar{S} &= \sqrt{\bar{\hat{q}}_S - \ln 2 \pi \bar{\hat{q}}_S + 2 \ln t},
    \label{S_selfcalib_app}
\end{align}
for large $\bar{\hat{q}}_S$. Note for sufficiently large $\bar{\hat{q}}_S$, $\bar{S} \approx \sqrt{\bar{\hat{q}}_S}$.
Eqs.~\ref{tlnNsc_app}, \ref{qap_app}, \ref{sidak_selfcalib_app}, and \ref{S_selfcalib_app} correspond to Eqs.~\ref{tlnNsc}, \ref{qap}, \ref{sidak_selfcalib}, and \ref{S_selfcalib}, quoted in the main text.


\bsp	
\label{lastpage}
\end{document}